**Participation in the Johari-Goldstein Process – Molecular Liquids versus Polymers**

D. Fragiadakis and C.M. Roland

Naval Research Laboratory, Chemistry Division, Washington DC 20375-5342

*(March 23, 2017)*

ABSTRACT We show using molecular dynamics simulations that simple diatomic molecules in the glassy state exhibit only limited participation in the Johari-Goldstein (JG) relaxation process. That is, with sufficient cooling local reorientations are essentially frozen for some molecules, while others continue to change their orientation significantly. Thus, the "islands of mobility" concept is valid for these molecular glass-formers; only near the glass transition temperature does every molecule undergo the JG process. In contrast, for a linear polymer this dichotomy in the distribution of JG relaxation strengths is absent – if any segments are changing their local orientation, all segments are.

The purpose of this note is to describe results showing that the limited participation in the Johari-Goldstein (JG) process reported for glassy polymers [1,2] and mixtures [3,4] is actually a salient property of molecular liquids; that is, notwithstanding suggestions to the contrary in a recent paper in this journal [5], a significant fraction of molecules in the glassy state may not change their orientation even while others are reorienting over significant angular ranges. Such dichotomy in the dynamics conforms to the "islands of mobility" description of the JG relaxation process [6]. The JG secondary relaxation is a universal feature of glass-forming liquids and polymers [7], and although it involves motion of all atoms in the molecule or polymer repeat unit, its limited amplitude results in faster dynamics than for structural relaxation. This leads to the idea that the JG process is the temporal precursor to, and thus fundamentally the cause of vitrification [8]. One implication of such a connection is that the well-established dynamic heterogeneity of structural relaxation may also be a property of the secondary JG process, despite the more restricted length scale of the latter. In previous work we found that the fourth-order dynamic susceptibility, $\chi_4(t)$, a measure of dynamic correlation, was non-zero for the JG process, attaining values as high as one-half that of $\chi_4(t)$ for structural relaxation. This indicates a significant degree of correlation.

Dynamic heterogeneity, a consequence of the reciprocal influences of molecular motions in condensed matter, is usually manifested both in spatial correlation, as reflected in $\chi_4(t)$, and in a distribution of relaxation rates (non-Debye behavior). The topic of interest herein is whether the distribution of rates in dynamically heterogeneous systems may extend to zero; that is, can some species (molecule or chain segment) not participate in the dynamics for conditions at which the mean relaxation time is finite. This behavior is known for structural relaxation, with an increasing cessation of translational and rotational mobility ultimately giving rise to glass formation. The situation is less clear for the JG relaxation. It presumes a lack of dynamic exchange among molecules over the JG time scale, as has been shown in molecular dynamic simulations (mds) of diatomic molecules [9,10]. However, in experiments on real molecular liquids, none have been identified in which not all molecules are involved in the JG process [5]. Since sufficiently deep within the glassy state even the local reorientations associated with JG motion cease, the question is merely whether there are state points at which some molecules retain fixed orientation while others are changing their angular position. Of course, the absence of known examples does not negate their possibility; accordingly, we undertook mds of two materials to determine if the distribution of JG relaxation strengths in the glassy state ever extends to zero. A further aim is ascertaining whether the JG process in molecular liquids and polymers is fundamentally different in this regard.

Simulations were carried out using the RUMD simulation software [11]. We studied two systems, we believe are the simplest molecular [9,10,12] and polymeric [13] systems that show a JG relaxation in MD simulations with the characteristics observed experimentally: (i) an asymmetric diatomic molecule having fixed bond length, and (ii) a linear polymer with bond angles constrained to 120 degrees. Atoms not connected covalently interact via the Lennard-Jones potential

$$U_{ij}(r) = 4\varepsilon_{ij}\left[\left(\frac{\sigma_{ij}}{r}\right)^{12} - \left(\frac{\sigma_{ij}}{r}\right)^{6}\right] \tag{1}$$

where *r* is the distance between atoms i and j. For the molecular system (1000 diatomic molecules) the energy and size parameters $\varepsilon_{ij}$ and $\sigma_{ij}$, were chosen based on the Kob-Andersen liquid [14], adjusted to give well separated JG and $\alpha$ dispersions (see ref. [9] for details). The bond length was maintained constant at $l_0$ = 0.45 using a constraint force algorithm [15]. For the polymer (20 linear chains with 100 monomers per chain) a single type of monomer was used with $\epsilon = \sigma = 1$. Bond lengths and bond angles were maintained approximately fixed (within a few percent) at $l_0 = 0.5$ and $\theta_0 = 120°$, respectively, using harmonic bond and bond angle potentials

$$U_{bond}(r) = \frac{1}{2}k\left(r - l_0\right)^2,$$

$$U_{angle}(\theta) = \frac{1}{2}k_\theta\left(\cos\theta - \cos\theta_0\right)^2$$

with a large constant $k = k_\theta = 3000$.

The systems are equilibrated at a temperature well above the glass transition ($T_{eq}$ = 1.0 for the molecular liquid and 5.0 for the polymer), then quenched to a temperature well below the glass transition; that is, a temperature at which the α relaxation is much slower than the simulation timescale and the system is out of equilibrium (T=0.33 for the molecular system, 2.0 for the polymer). This glassy system was equilibrated until no significant aging (drift in pressure, potential energy or dynamic correlation functions) is observed during the simulation timescale.

Displayed in Figure 1 for twenty representative diatomic molecules is the first-order rotational correlation function in the glassy state

$$C(t) = \langle \cos\theta(t) \rangle \tag{2}$$

where $\theta$ is the angular change in the molecular axis. Initially for all molecules there is a ~10% decay due to vibrations (cage rattling). The vibrational time of picoseconds for real materials calibrates our simulation time scale. Beyond *ca.* $t$ = 0.2, another decay ensues, corresponding to the JG relaxation. Diverse behavior is observed, with the individual *C(t)* exhibiting very different relaxation strengths, $\Delta_{JG}$ (the magnitude of the second decay in C(t)). The mean C(t) reaches a plateau of 0.45 after about t = $10^5$. The simulation was continued for a duration *ca.* 500 times the average JG relaxation time, $\tau_{JG}$ (but several decades shorter than the time for the onset of structural relaxation). Over this time span, some molecules lose nearly all correlation with their initial angular position, while others show no change. This dichotomy is seen clearly in the distribution function for the relaxation strengths (Fig. 1 inset). There is a continuous distribution of strengths having a maximum at a value larger than $\overline{\Delta}_{JG}$ and extending to $\Delta_{JG}$=1. Simultaneously for a significant fraction of molecules, $\Delta_{JG}$ is very small, and the distribution shows an upturn towards $\Delta_{JG}$=0.

Figure 1 also shows the same data for twenty linear polymer chains. It is obvious that the spread of relaxation strengths attained beyond $\tau_{JG}$ is much narrower than for the molecular glass. Moreover, the smallest degree of relaxation is still significantly more than that ascribable to vibrations; that is, every chain segment is undergoing angular changes associated with the JG process. The distribution function for these relaxation strengths, displayed in the figure inset, is narrow and centered on $\overline{\Delta}_{JG}$. The polymer chains have structural diversity due to the chain ends that is not present in the diatomic molecules. In Figure 2 we have included the mean *C(t)* for the chain ends. While the JG relaxation of these segments is about one decade faster than the average of all segments, there is no difference in the relaxation strength of segments at the chain terminus versus interior segments.

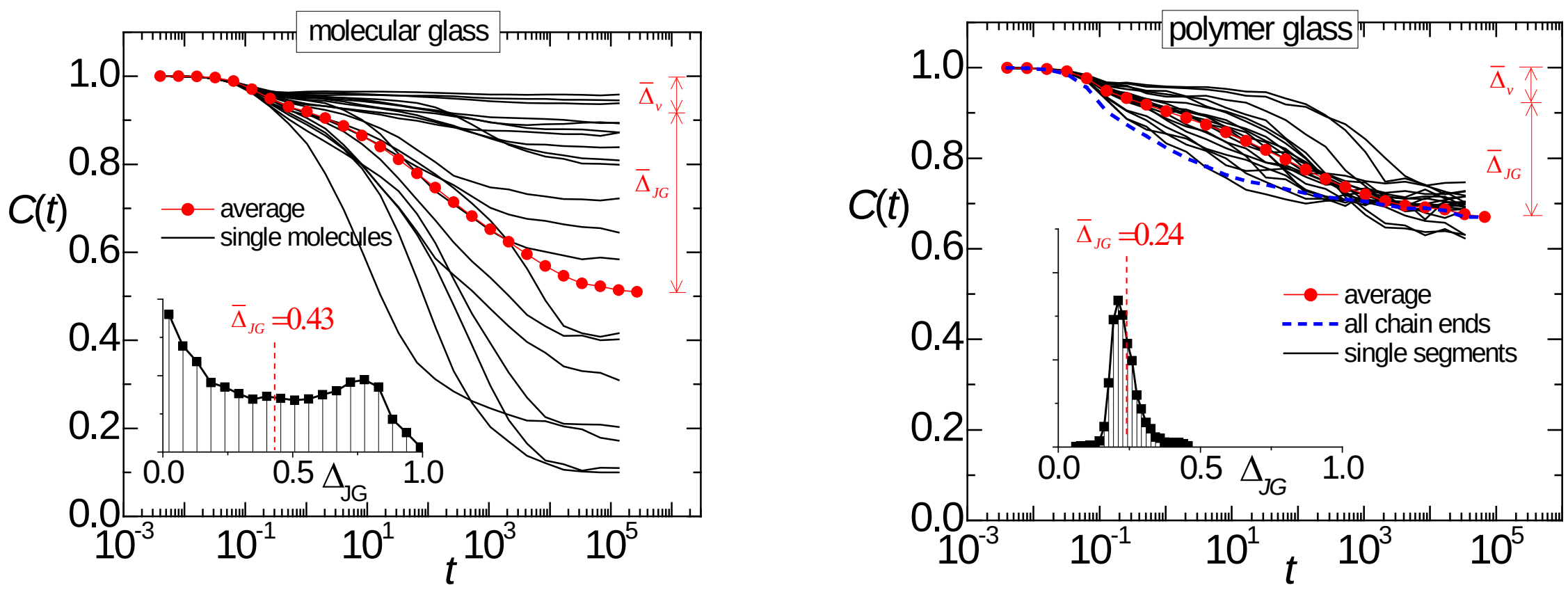

**Figure 1.** (left) First-order rotational correlation function of representative single molecules (thin lines) and C(t) averaged over all molecules (line with circles). (right) C(t) for representative single chain segments (thin lines), and the average C(t) for all segments (line with circles) and for all segments at chain ends only (dashed line). The insets show the distribution of $\Delta_{JG}$, which is the height of the step corresponding to the JG relaxation.

We have previously shown that for the asymmetric diatomic molecules, the JG relaxation comprises mainly large angular jumps. These can be seen in the self-part of the angular van Hove function [16,17]

$$G_s(\theta,t)=\frac{2}{N\sin\theta}\sum_{i=1}^{N}\delta\left[\theta-\theta_i(t)\right].$$

The probability that the molecular axis (or segment axis, for polymers) lies on an angle between $\theta$ and $\theta+d\theta$ at time $t$ for an initial orientation $\theta=0$ is given by $\frac{1}{s}G_s(\theta,t)\sin\theta d\theta$. Figure 2 shows $G_s(\vartheta, t)$ for the two simulated systems. The broad peaks at θ~100° and 150° for the molecular glass are a signature of the JG relaxation; they appear at the beginning of the JG regime, and grow as the relaxation progresses. The polymer exhibits similar behavior; however, there is only a single peak, around 60°. The peak which is significantly narrower than for the diatomic, reflecting the greater homogeneity of the JG relaxation.

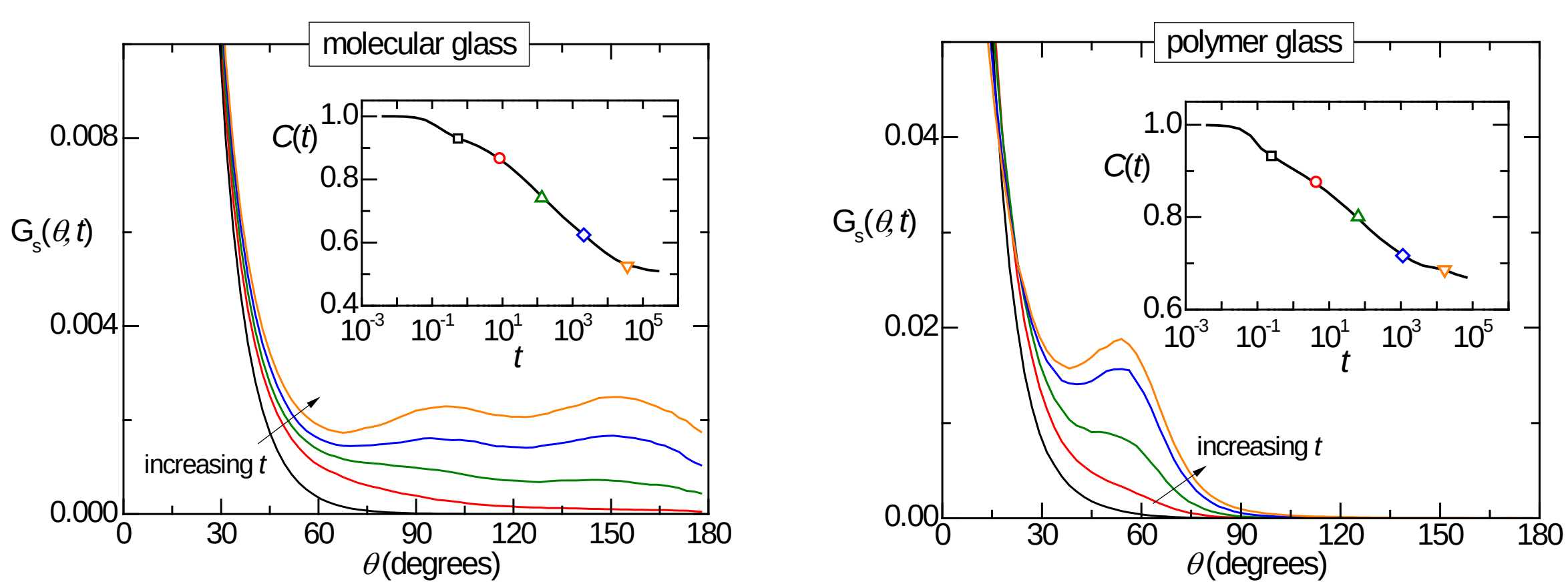

**Figure 2.** Self part of the angular Van Hove function at times within the JG regime for (left) molecular glass and (right) polymer glass. Each curve corresponds to one of the points in C(t) plotted in the insets.

In summary, we have shown from mds that the distribution of JG relaxation strengths for diatomic molecules can extend to zero; that is, some molecules remain essentially stationary despite rotational motion of neighboring molecules. Molecules whose orientation almost completely decorrelates on the JG timescale coexist with molecules that do not change their orientation over times at least 500 times longer than the mean JG relaxation time. These results are in accord with the islands of mobility description of JG secondary relaxations [6]. Of course, closer to the glass transition, the mean JG relaxation time becomes very short, and all molecules reorient. This latter circumstance is the one commonly investigated in multi-dimensional NMR experiments [1,2]. For polymers we did not observe the islands of mobility scenario. At temperatures sufficiently high for the JG process to proceed, all repeat units participate. The inference is that segments cannot maintain a fixed orientation when directly bonded to segments undergoing angular changes. Different behavior may obtain for polymers having pendant moieties from which the JG dynamics may originate, such as acrylate polymers [1,2,18]. Of course the behaviors observed herein were for mds covering the entire temporal range of the JG process, made possible by judicious selection of the structural parameters of our simulated materials. In real systems observation of the broad JG relaxation can be limited by overlap with the structural relaxation dispersion. Additionally, mds uniquely provides information about each molecule or repeat unit, whereas even for multi-dimensional NMR the ability to discriminate among relaxing entities is more limited.

**ACKNOWLEDGEMENT**

This work was supported by the Office of Naval Research.